\documentclass[prd,twocolumn,superscriptaddress,letterpaper,nofootinbib,nopreprintnumbers,noshowpacs]{revtex4-1}

\usepackage{amsmath}
\usepackage{amssymb}
\usepackage{graphicx}
\usepackage[caption=false]{subfig}
\usepackage{dcolumn}
\usepackage{bm} 
\usepackage{hyperref}
\usepackage{epsfig} 
\usepackage{color} 
\usepackage[utf8]{inputenc}
\usepackage{comment}
\usepackage[all]{hypcap}

\newcommand{\BBHDetections}{5}
\newcommand{\BNSDetections}{1}

\newcommand{\compact}{ultracompact}

\begin{document}


\title{Methods for the detection of gravitational waves from sub-solar mass \compact{} binaries}

\author{Ryan Magee}
\affiliation{Institute for Gravitation and the Cosmos, The Pennsylvania
State University, University Park, PA 16802, USA\\}
\affiliation{Department of Physics, The Pennsylvania State University, University
Park, PA 16802, USA\\}
\author{Anne-Sylvie Deutsch}
\affiliation{Institute for Gravitation and the Cosmos, The Pennsylvania
State University, University Park, PA 16802, USA\\}
\affiliation{Department of Physics, The Pennsylvania State University, University
Park, PA 16802, USA\\}
\author{Phoebe McClincy}
\affiliation{Department of Physics, The Pennsylvania State University, University
Park, PA 16802, USA\\}
\author{Chad Hanna}
\affiliation{Institute for Gravitation and the Cosmos, The Pennsylvania
State University, University Park, PA 16802, USA\\}
\affiliation{Department of Physics, The Pennsylvania State University, University
Park, PA 16802, USA\\}
\affiliation{Department of Astronomy and
Astrophysics, The Pennsylvania State University, University Park, PA 16802,
USA\\}
\author{Christian Horst}
\affiliation{University of Wisconsin-Milwaukee, Milwaukee, Wisconsin 53201, USA\\}
\author{Duncan Meacher}
\affiliation{Institute for Gravitation and the Cosmos, The Pennsylvania
State University, University Park, PA 16802, USA\\} 
\affiliation{Department of Physics, The Pennsylvania State University, University
Park, PA 16802, USA\\}
\affiliation{University of Wisconsin-Milwaukee, Milwaukee, Wisconsin 53201, USA\\}
\author{Cody Messick}
\affiliation{Institute for Gravitation and the Cosmos, The Pennsylvania
State University, University Park, PA 16802, USA\\} 
\affiliation{Department of Physics, The Pennsylvania State University, University
Park, PA 16802, USA\\}
\author{Sarah Shandera}
\affiliation{Institute for Gravitation and the Cosmos, The Pennsylvania
State University, University Park, PA 16802, USA\\} 
\affiliation{Department of Physics, The Pennsylvania State University, University
Park, PA 16802, USA\\}
\author{Madeline Wade}
\affiliation{Kenyon College, Gambier, Ohio 43022, USA\\}

\date{\today}
\begin{abstract}

We describe detection methods for extensions of gravitational wave searches to
sub-solar mass compact binaries.  Sub-solar mass searches were previously
carried out using Initial LIGO, and Advanced LIGO boasts a detection volume
approximately $1000$ times larger than Initial LIGO at design sensitivity. Low
mass compact binary searches present computational difficulties, and we suggest
a way to rein in the increased computational cost while retaining a sensitivity
much greater than previous searches.  Sub-solar mass compact objects are of
particular interest because they are not expected to form astrophysically. If
detected they could be evidence of primordial black holes (PBH). We consider a
particular model of PBH binary formation that would allow LIGO/Virgo to place
constraints on this population within the context of dark matter, and we
demonstrate how to obtain conservative bounds for the upper limit on the dark
matter fraction.

\end{abstract}


\maketitle

\section{Introduction} \label{intro}

Advanced LIGO~\cite{2015CQGra..32g4001L} and Advanced Virgo's~\cite{2015CQGra..32b4001A}
detections of gravitational waves from compact binary coalescences (CBC) have
ushered in the dawn of gravitational wave astronomy. To date, there have been
\BBHDetections{} detections of binary black hole
mergers~\cite{Abbott:2016nmj,Abbott:2017vtc,Abbott:2017gyy, Abbott:2017oio} and
\BNSDetections{} detection of a binary neutron star
system~\cite{TheLIGOScientific:2017qsa}, each of which has expanded our
knowledge of the properties and populations of compact objects in our universe.
Advanced LIGO and Advanced Virgo's success in detecting traditional sources of
gravitational waves suggest that ground based interferometers could be powerful
new tools in observing the dark universe. We describe considerations for
extensions of traditional compact binary searches to the sub-solar mass regime,
and provide motivation for these searches in the context of dark matter. In
particular, we consider a model where a uniform distribution of monochromatic
primordial black holes (PBH) make up a fraction of the dark matter. We examine
the model's robustness and demonstrate how it can place constraints on the
abundance of PBHs for different sub-solar mass populations.

\section{Analysis Techniques}

LIGO compact binary searches rely on matched filtering to extract candidate
signals from the noise by correlating known gravitational waveforms with the
data. Compact binary searches currently require
$\mathcal{O}(10^5)-\mathcal{O}(10^6)$ templates to adequately recover arbitrary
signals placed in the parameter spaces considered thus far (binary systems with a total mass of
$2M_\odot - 600 M_\odot$~\cite{TheLIGOScientific:2016pea,
Abbott:2017iws}). The addition of fully precessing waveforms in future
observing runs could increase this by yet another factor of $10$, though for
now this remains computationally infeasible. 

The difficulty of CBC searches scales with both the number and length of the
waveforms used as matched filter templates, which could present a problem for
sub-solar mass searches.  Here we focus on the effect of the number of templates in the template bank which is
expected to scale (roughly) as:
\begin{align} \label{eq:scaling} N \propto m_{\text{min}}^{-8/3} f_{\text{min}}^{-8/3} \end{align}%
where $m_{\text{min}}$ is the minimum mass included in the search and
$f_{\text{min}}$ denotes the starting frequency of the template
waveforms~\cite{Owen:1998dk}. Previous Advanced LIGO
searches have searched for binaries with components as light as $1M_\odot$
~\cite{TheLIGOScientific:2016pea,Abbott:2016ymx}; extending these searches to
lower masses could easily lead to a $10 - 100$ time increase in difficulty
compared to offline analyses in Advanced LIGO's first observing run.  Below we
propose increasing $f_{\text{min}}$ to mitigate the increased computational
costs associated with low mass extensions of compact binary searches, and we
calculate the expected loss in sensitivity that this brings.

\subsection{Estimates of sensitivity} \label{sensitivity}

Second-generation ground-based gravitational wave detectors such as Advanced
LIGO and Advanced Virgo are sensitive over a broad range of frequencies ($\sim
10 - 10\thinspace000\, \text{Hz}$) but they are most sensitive near $100 \,
\text{Hz}$~\cite{Martynov:2016fzi}. Compact binary pipelines exploit this
sensitivity and typically analyze a subset of the total bandwidth. In Advanced
LIGO's first observing run, frequencies spanning $10-2048 \, \text{Hz}$ were
analyzed~\cite{Messick:2016aqy}. This is an excellent approximation for
standard CBC searches; the majority of the signal-to-noise ratio (SNR) is
accumulated at lower frequencies and very little sensitivity is lost by cutting
the analysis at $2048 \text{Hz}$. This is an even better approximation for
sub-solar mass compact binaries since the frequency evolution of a binary goes
as~\cite{Cutler:1994ys}:
\begin{align} \dot{f} \propto \mathcal{M}^{5/3} f^{11/3} \end{align}
where 
\begin{align} \mathcal{M} = \frac{(m_1 m_2)^{3/5}}{(m_1 + m_2)^{1/5}}
\end{align} is the chirp mass of the system. Sub-solar mass systems therefore are not
only long lived, but also spend a long time in LIGO's most sensitive band
compared to heavier binaries. This suggests that it may be possible to analyze
an even more reduced frequency band than previous searches while retaining a
significant amount of SNR.

Since orbital decay is slow for sub-solar mass \compact{} binaries, inspiral
only waveforms are a very good approximation of the signal received on earth.
The amplitude of the waveform can be written as~\cite{Abadie:2010cg}:
\begin{align} \left | \tilde{h}(f)\right |  = \frac{1}{D} \left(\frac{5 \pi}{24 c^3}\right)^{1/2} (G\mathcal{M})^{5/6} (\pi f)^{-7/6} \end{align}
and the average recovered signal to noise ratio is given by:
\begin{align} \langle \rho \rangle = \sqrt{4 \int_{f_{\text{min}}}^{f_{\text{max}}} \frac{
\left | \tilde{h}(f) \right |^2}{S_n(f)} df} \end{align}
where $S_n(f)$ denotes the single sided power spectral density, informally
referred to as the ``noise curve''. $f_{\text{min}}$ is determined by either the low
frequency noise floor or the starting frequency of the template waveform
(whichever is greater) and $f_{\text{max}}$ is determined by the frequency of the
innermost stable circular orbit ($f_{\text{ISCO}}$) or the ending frequency of
the template waveform (whichever is less) where $f_{\text{ISCO}}$ is defined as:
\begin{align} f_{\text{ISCO}} = \frac{c^3}{6\sqrt{6}\pi G M_{\text{total}}}   \end{align} 
For a $1 M_\odot - 1 M_\odot$ binary, $f_{\text{ISCO}} \approx 2200\, \text{Hz}$. The
frequency monotonically increases for lighter total mass systems; for a
sub-solar mass search, $f_{\text{max}}$ is determined by the bandwidth of the
template waveforms.

\begin{figure}
\includegraphics[width=\columnwidth]{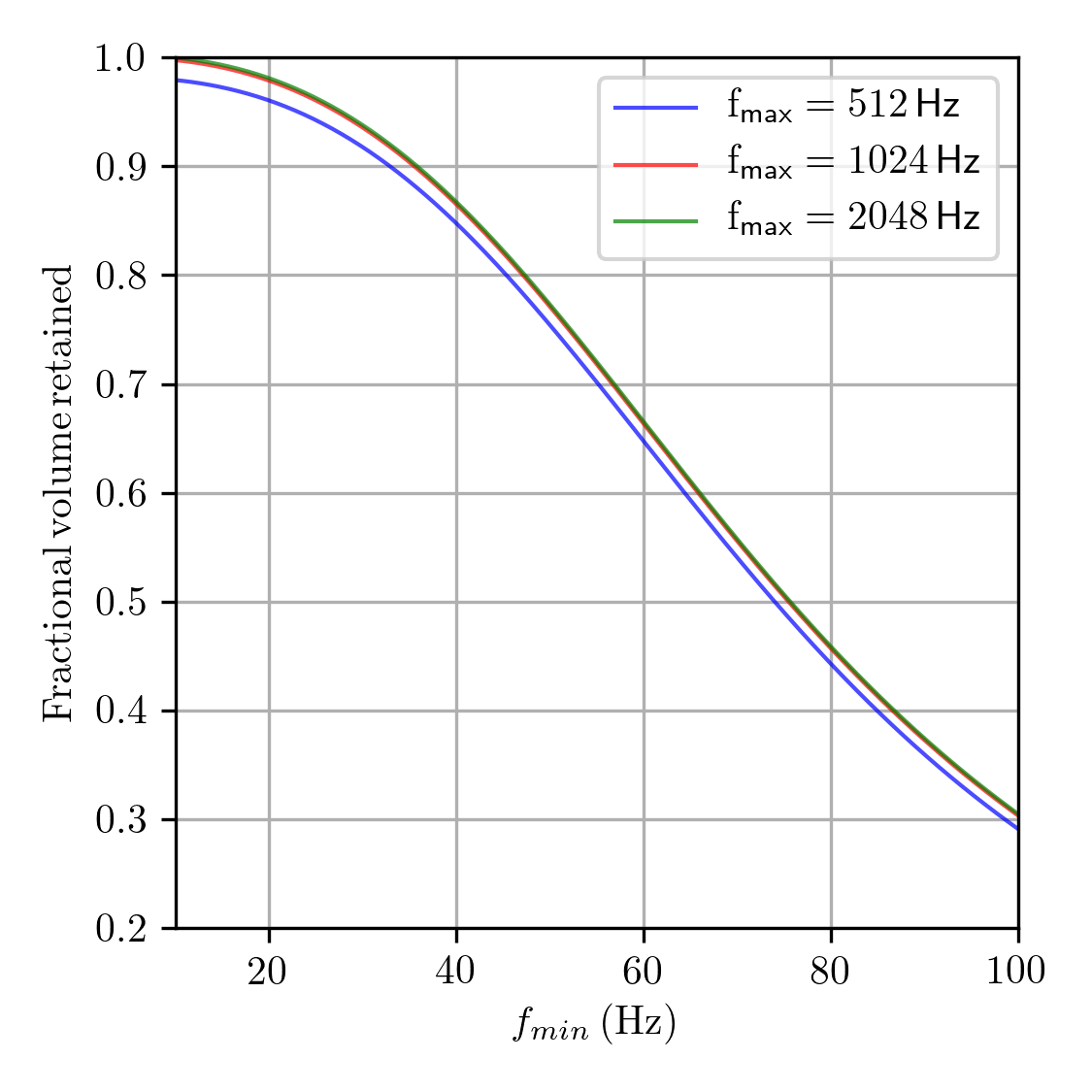}
\caption{\label{fig:snrloss} The fractional volume retained for various values
of $f_{\text{max}}$ and as a function of $f_{\text{min}}$. The green, red, and
blue lines correspond to upper cut-off frequencies of $2048$,
$1024$, and $512$ Hz respectively. Note that there is very little difference
between the various $f_{\text{max}}$ values; this is because there is more than
an order of magnitude more noise at these frequencies than the $\sim 100 \,
\text{Hz}$ region and very little SNR is accumulated there. All values are
measured relative to the band $f \in (10 \, \text{Hz}, \, 2048 \, \text{Hz})$.
} \end{figure}

We can substitute the waveform amplitude into the equation for SNR and
rearrange to find the horizon distance for a given $\langle \rho \rangle$ (or equivalently, the SNR recovered at some fiducial distance):
\begin{align} \label{eq:horizon} D_{max} \propto \frac{1}{\langle \rho \rangle} \mathcal{M}^{5/6} \sqrt{4\int_{f_{\text{min}}}^{f_{\text{max}}} \frac{f^{-7/3}}{S_n(f)}df} \end{align}
which is dependent on the noise curves, the chirp mass of the binary, and the
frequency band of the analysis. This allows us to compare LIGO's
sensitivity for frequency bands that do not encompass the full sensitive range.
We choose the $f \in (10 \text{Hz}, \, 2048 \text{Hz})$ band as a point of
comparison. The fraction of SNR retained is then:
\begin{align} f_{\text{SNR}} = \frac{D(f_{\text{min}}, \, f_{\text{max}})}{D(10 \, \text{Hz}, \, 2048 \, \text{Hz})} \end{align}
Note that this fractional reduction is independent of the mass of the binary. This
presents an important trade off in sub-solar mass searches: increasing
$f_{\text{min}}$ drives the difficulty of a search down, but it also causes the search
to lose sensitivity. This drop in SNR is equivalent to a fractional decrease in
LIGO's average range, which means that the observed volume (and therefore the
expected number of detections at a given chirp mass) is smaller by a factor of
$f_{\text{SNR}}^3$. Thus even a $3\%$ loss in SNR would represent a detection
volume nearly $10\%$ smaller. The sensitive volume retained as a function of
$f_{\text{min}}$ and $f_{\text{max}}$ is shown in Fig.~\ref{fig:snrloss}. 

\subsection{Sensitive distance}
 
Initial LIGO previously carried out searches for compact binaries with
components as light as $0.2 M_\odot$~\cite{Abbott:2005pf}. Using the relations
outlined above and the fact that current Advanced LIGO searches extend to $1
M_\odot$ and $f_{\text{min}} = 10 \, \text{Hz}$, we can estimate the reduction
in frequency band and sensitivity required to keep the cost of a sub-solar mass
search comparable to current Advanced LIGO searches. Equation~\ref{eq:scaling}
shows that we expect similar scalings in both $m_{\text{min}}$ and
$f_{\text{min}}$. Thus if we decrease the lower mass bound of previous Advanced
LIGO searches by a factor of 5, we need to \emph{increase} $f_{\text{min}}$ by
a factor of 5 as well to keep the number of templates approximately constant.
We estimate that in order to modify current searches to extend down to this
mass we would need to increase $f_{\text{min}}$ to $\sim 50 \text{Hz}$. This
amounts to a loss of $10\%$ in SNR and range, and therefore a loss of $\sim
30\%$ in volume and detection rate.  Even with this loss in sensitivity, LIGO
would remain incredibly sensitive to sub-solar mass \compact{} binaries. 

To emphasize LIGO's improvement in sensitivity even with this reduction in
recovered SNR, consider the most recent search for sub-solar mass compact
objects carried out in Initial LIGO's third and fourth science runs.  The
lowest mass binary considered in this search remained visible at a range of
$\sim 4 \, \text{Mpc}$~\cite{Abbott:2007xi}. The estimate outlined here suggests
that at the same mass and over a reduced frequency band, Advanced LIGO has a
range of $\sim 45 \, \text{Mpc}$ which corresponds to a sensitive volume more than
$1000$ times greater. The massive increase in physical volume accessible by
Advanced LIGO vastly outweighs any loss in sensitivity due to a moderately
reduced frequency band (provided analyzable time remains approximately the
same).  

\begin{figure} \includegraphics[width=\columnwidth]{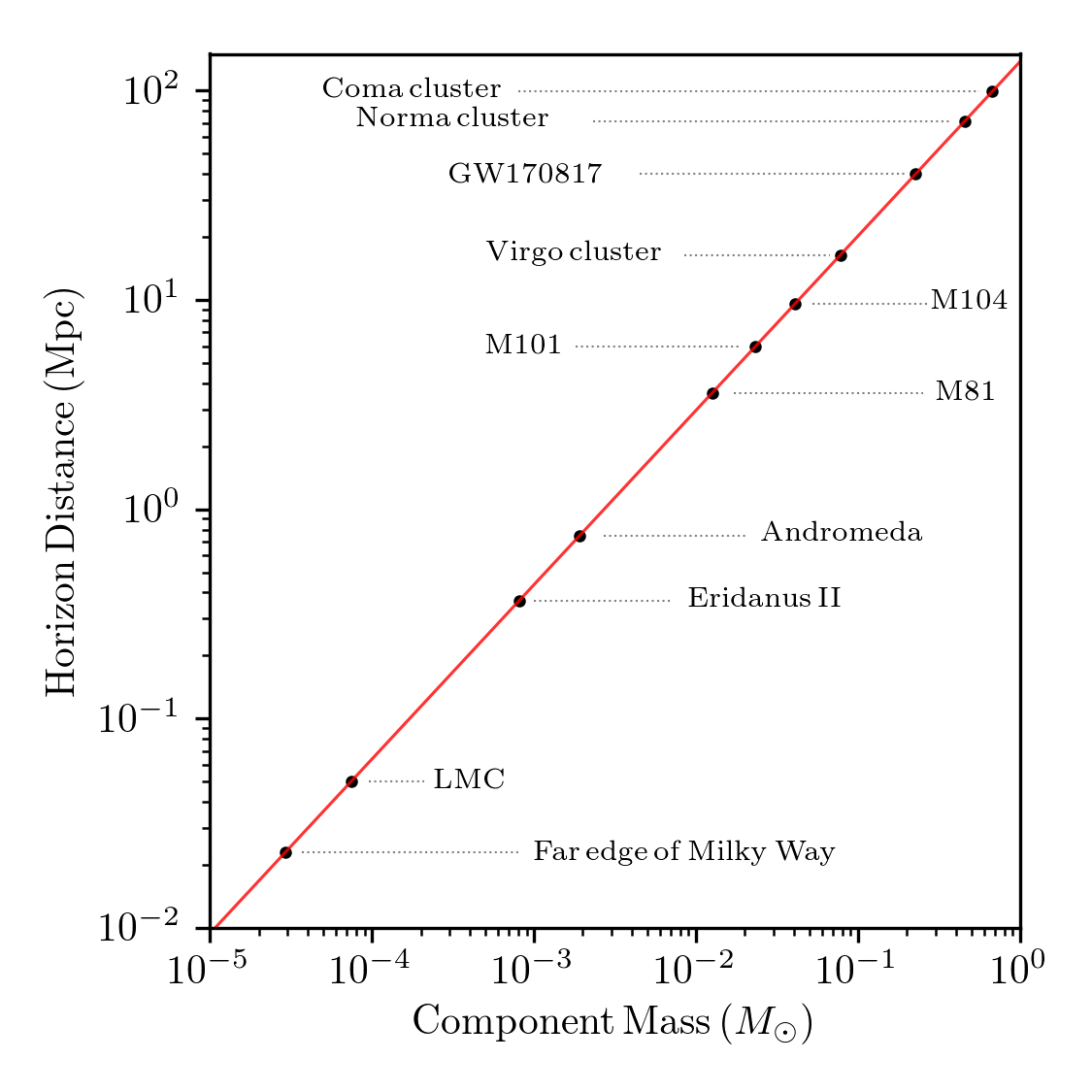}
\caption{\label{fig:horizon} The distance to an optimally oriented, equal mass
binary shown as a function of the component mass. LIGO remains sensitive to
$\mathcal{O}( 10^{-5} - 10^{-5} M_\odot)$ binaries at extra-galactic distances.
This plot assumes $f_{\text{min}} = 10\,  \text{Hz}$ and $f_{\text{max}} =
2048\, \text{Hz}$ and therefore represents an optimistic view of horizon
distance and ignores search difficulty. Astrophysical galaxies, groups, and
clusters are included as a reference for cosmological distances. Several objects
previously considered as observational candidates for the abundance of dark
matter (Eridanus II, LMC/SMC, Segue I) are well within LIGO's range at low
masses. Approximate distances taken
from~\cite{TheLIGOScientific:2017qsa,Mei:2007xs,Pietrzynski:2013gia,2010A&A...509A..70V,Carter:2008jg,2011ApJ...743..176G,Mutabazi:2014esa}.
The noise curve used to approximate O1 sensitivity is ``Early high/Mid low'' from~\cite{Martynov:2016fzi}.  }
\end{figure}

\subsection{Approximation of the merger rate for null-results} \label{rates_estimate}

In Equation~\ref{eq:horizon} we defined the horizon distance of the detector.
This represents the maximum distance for which an optimally located and oriented
source would be recovered with some $\langle \rho \rangle$. In general,
however, detectors will measure a weaker response to a gravitational wave depending
on the location and orientation of the binary. This reduction is described by
the antenna patterns, $F_+$ and $F_\times$, which always take values less than
or equal to 1 and are related to the signal observed on earth through:
\begin{align} h = F_+ h_+ + F_\times h_\times \end{align}
Averaging the detector response over both location and orientation of the
binary reduces the the strain recovered (and therefore the distance to a binary
with some fiducial $\langle \rho \rangle$) by a factor of
2.26~\cite{Finn:1992xs,Abadie:2010cf,Sutton}. This can be used to define the
average range of the detector as
\begin{align} D_{avg} = \frac{D_{max}}{2.26} \end{align}

The average sensitive distance allows us to approximate limits on the
coalescence rate from null results for a general gravitational wave search. The
loudest event statistic formalism~\cite{Biswas:2007ni} states that we can
constrain the binary merger rate for a specific mass bin, $i$, to $90\%$
confidence with:
\begin{align}\mathcal{R}_{90, i} = \frac{2.3}{\langle VT \rangle_i}\end{align}
We can estimate the sensitive volume-time for a particular observing run using
the earlier range approximation. 
\begin{align} \langle VT \rangle_i = \tfrac{4}{3} \pi D_{avg, i}^3 T \end{align}
where $T$ is the analyzable live-time of the two detectors. This method
provides an excellent approximation of the sensitive 4-volume. The remaining plots in
this paper use this procedure to estimate LIGO rates and LIGO sensitivity in
the sub-solar mass region.

\subsection{Non-spinning waveforms}

While reducing the frequency band is one way to mitigate the increased
computational cost of sub-solar mass searches, non-spinning waveforms also
offer an easy way to reduce the difficulty by potentially $1-2$ orders of
magnitude. There are some theoretical justifications for non-spinning searches: some
models predict sub-solar mass black holes to be predominately slowly
spinning~\cite{Chiba:2017rvs}, and LIGO's previous detections have been
consistent with low $\chi_{\text{eff}}$ binaries.  Regardless, a completely
non-spinning binary is clearly a non-physical assumption. The efficacy of using
non-spinning waveforms to recover spinning waveforms has been examined
before~\cite{Cho:2017yaj,Capano:2016dsf,Canton:2014ena}. In particular,
~\cite{Cho:2017yaj} examined neutron star systems and found that non-spinning
templates recovered aligned spin binary neutron stars to the desired level only
for $-0.2 \lesssim \chi_{\text{eff}} \lesssim 0$. 

We performed a similar test on a population of $.5 M_\odot - .5 M_\odot$ binary
black holes. We created a non-spinning template bank covering component masses
$ m_i \in (0.3 M_\odot, \, 0.7 M_\odot) $ using TaylorF2
waveforms~\cite{Ajith:2012mn,Capano:2016dsf}. We then injected
$10\thinspace000$ spinning signals that were purely aligned or anti-aligned
with the orbital angular momentum and had dimensionless spin values of $\left |
\chi_i \right | < 0.5$ into fake data. We then calculated the overlap between our
non-spinning template waveforms and the spinning signals. We find results
similar to those of~\cite{Cho:2017yaj}; at low spin, there is a large overlap
between the template waveforms and the injected, spinning signals. At higher
spins, however, the maximum overlap rapidly falls off, implying that LIGO would
miss a significant fraction of the signals with appreciable spin. In fact, we
find that the non-spinning bank used here recovers signals well provided
$\chi_{\text{eff}} > -.08$ or $\chi_{\text{eff}} < .02$. As $\chi_{\text{eff}}$
deviates form these values, the fraction of signals missed grows rapidly. A
spinning template bank is therefore necessary if sub-solar mass \compact{}
binaries are either born with appreciable spin components or accrete enough
matter to develop substantial spin. We are currently examining the effects of
spin on the computational cost of sub-solar mass CBC searches, as well as other
possible ways to mitigate the increased difficulty.

\begin{figure} \centering
\includegraphics[width=\columnwidth]{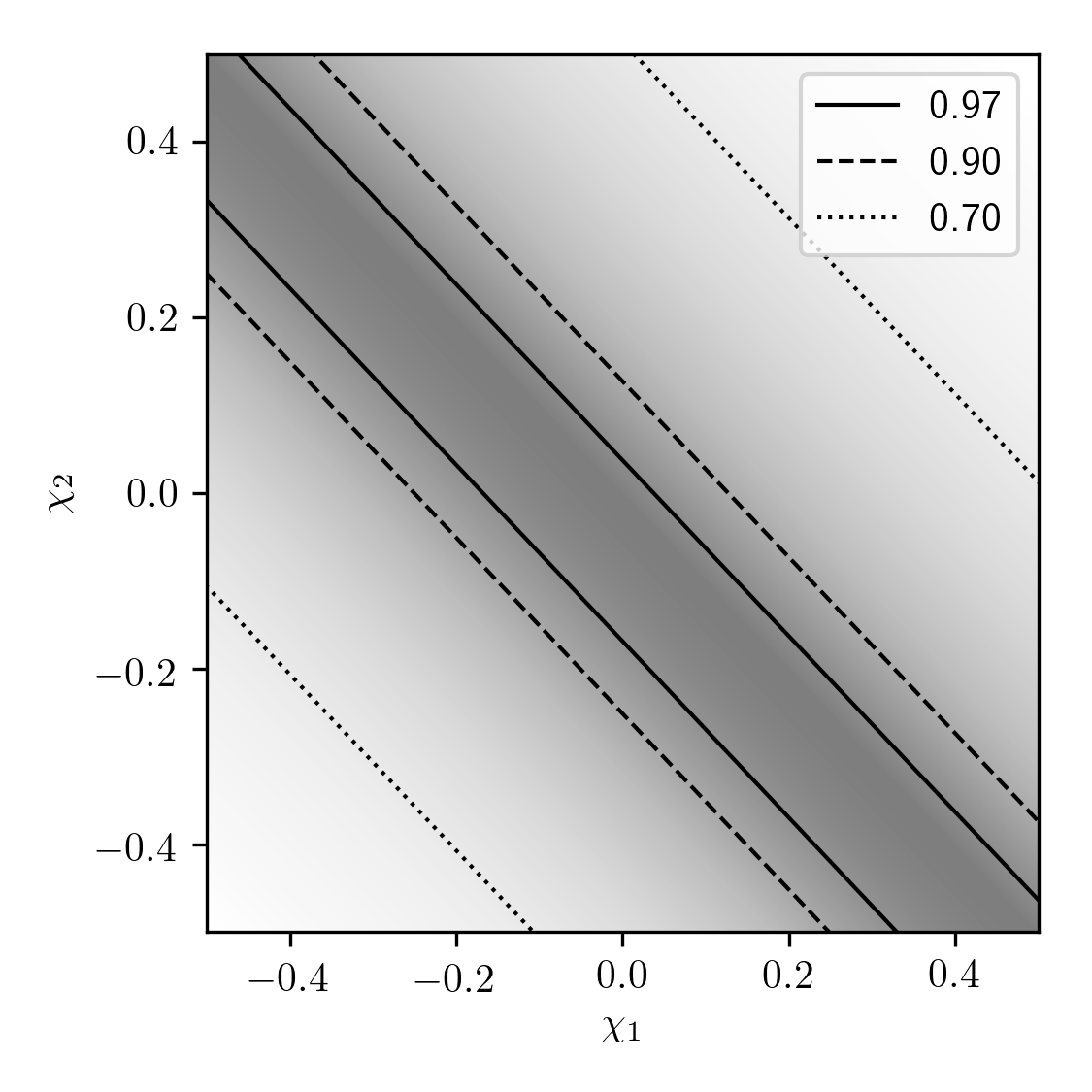}
\caption{\label{fig:spin} Recovery of spinning signals with a family of
non-spinning template waveforms. Shown in black are lines of constant fitting
factor (i.e. the maximum overlap between template waveforms and the injected
signals) with the value specified by the line type. The shading shows how the
fitting factor changes with the spin of the components in regions between the
contours.  While systems with $-.084 < \chi_{\text{eff}} < .019$ are recovered
well, the match between the two waveforms drops rapidly for $\chi_{\text{eff}}$
outside this range. The SNR is proportional to the fitting factor, so the loss
in SNR grows rapidly with total spin.} \end{figure}

\section{Potential constraints on primordial black hole abundance}

While there is a large population of compact objects below one solar mass, the
only objects compact enough for detection by LIGO are black holes and neutron
stars. Other compact objects begin to coalesce at too low of an orbital
frequency to produce gravitational waves in the sensitive band of ground-based
interferometers. Neither black holes nor neutron stars are expected to form
below one solar mass via known astrophysical mechanisms, though there are
models that propose alternative ways to form black holes at this
mass~\cite{Shandera:2018xkn,Kouvaris:2018wnh}.  It is interesting to consider
the possibility that sub-solar mass black holes are formed via primordial
processes and could be a component of the dark matter. In the event of either a
detection or null-result LIGO can provide estimates on the merger rate, so it
is therefore necessary to model the binary formation rate for primordial black
holes in order to connect LIGO with primordial populations. Here we describe
the sensitivity of one particular model to changes in input parameters, as well
as the response of constraints on the dark matter fraction, $f_{\text{PBH}}
\equiv \Omega_{\text{PBH}}/\Omega_{\text{DM}}$, to changes in merger rate
constraints that could be provided by LIGO. We motivate this model as a way to
provide a conservative limit on $f_{\text{PBH}}$.  

We consider a model of (initially) uniformly distributed, monochromatic black
holes formed in the early universe. A pair of nearest neighbor black holes will
start to decouple from the background cosmological expansion and form a binary
when the mean energy density in a volume encompassing the two exceeds the
background energy density. A third, closest black hole to the binary injects
angular momentum in the system by applying tidal forces, which ensures that the
two black holes will orbit rather than collide head-on.  The resulting
expression for the merger rate of primordial black hole binaries in the local
universe is given by:
\begin{equation} \text{event rate} = n_{\text{PBH}} \left. \frac{dP}{dt} \right|_{t=t_0}.  \end{equation}
where $dP$ is given by:
\begin{equation} dP = \begin{cases} \dfrac{3 \, f_{\text{PBH}}^{\tfrac{37}{8}}}{58}
\left[f_{\text{PBH}}^{-\tfrac{29}{8}} \left(\frac{t}{t_c}\right)^{\tfrac{3}{37}} -
\left(\frac{t}{t_c}\right)^{\tfrac{3}{8}} \right] \dfrac{dt}{t}, & \ t < t_c
\\[15pt] \dfrac{3 \, f_{\text{PBH}}^{\tfrac{37}{8}}}{58} \left[ f_{\text{PBH}}^{-\tfrac{29}{8}}
\left(\frac{t}{t_c}\right)^{-\tfrac{1}{7}} -
\left(\frac{t}{t_c}\right)^{\tfrac{3}{8}} \right] \dfrac{dt}{t}, & \ t \geq t_c
\end{cases} \end{equation}
and $n_{\text{PBH}}$ by:
\begin{equation} n_{\text{PBH}} = \frac{3H^2_0}{8 \pi G} \frac{\Omega_{\text{PBH}}}{M_{\text{PBH}}} \end{equation}
where
\begin{equation} t_c = Q \alpha^4 \beta^7 \bar{x}^4 f^{25/3} \end{equation}
and
\begin{equation}\bar{x} = \frac{1}{(1+z_{\text{eq}})} \left( n_{\text{PBH}} \right) ^{-1/3} \end{equation}
with $Q = 3 / 170 \, (GM_{\text{PBH}})^{-3}$, $G$ the gravitational constant,
$z_{\text{eq}}$ the redshift at matter-radiation equality, and $M_{\text{PBH}}$
the mass of each individual black hole in this population. $\alpha$ and $\beta$
are constants of $\mathcal{O}(1)$ that depend on the dynamics of binary
formation and are typically set to 1. This model has been extensively
studied~\cite{Ioka1998,Nakamura:1997sm,Sasaki:2016jop,Eroshenko:2016hmn,Wang:2016ana}.

This model provides a direct connection between LIGO and PBHs via an expected
merger rate which is solely a function of the age of the universe, $t_0$, given some $M_{\text{BH}}$ and
$f_{\text{PBH}}$. The merger rate is not analytically invertible, but if gravitational
wave observations provide a constraint on the merger rate for black holes of a
particular mass, then it can be numerically solved to obtain an upper limit on
$f_{\text{PBH}}$ for that mass bin. Similar procedures have been considered
before~\cite{Sasaki:2016jop,Wang:2016ana}. 

\begin{figure} \includegraphics[width=\columnwidth]{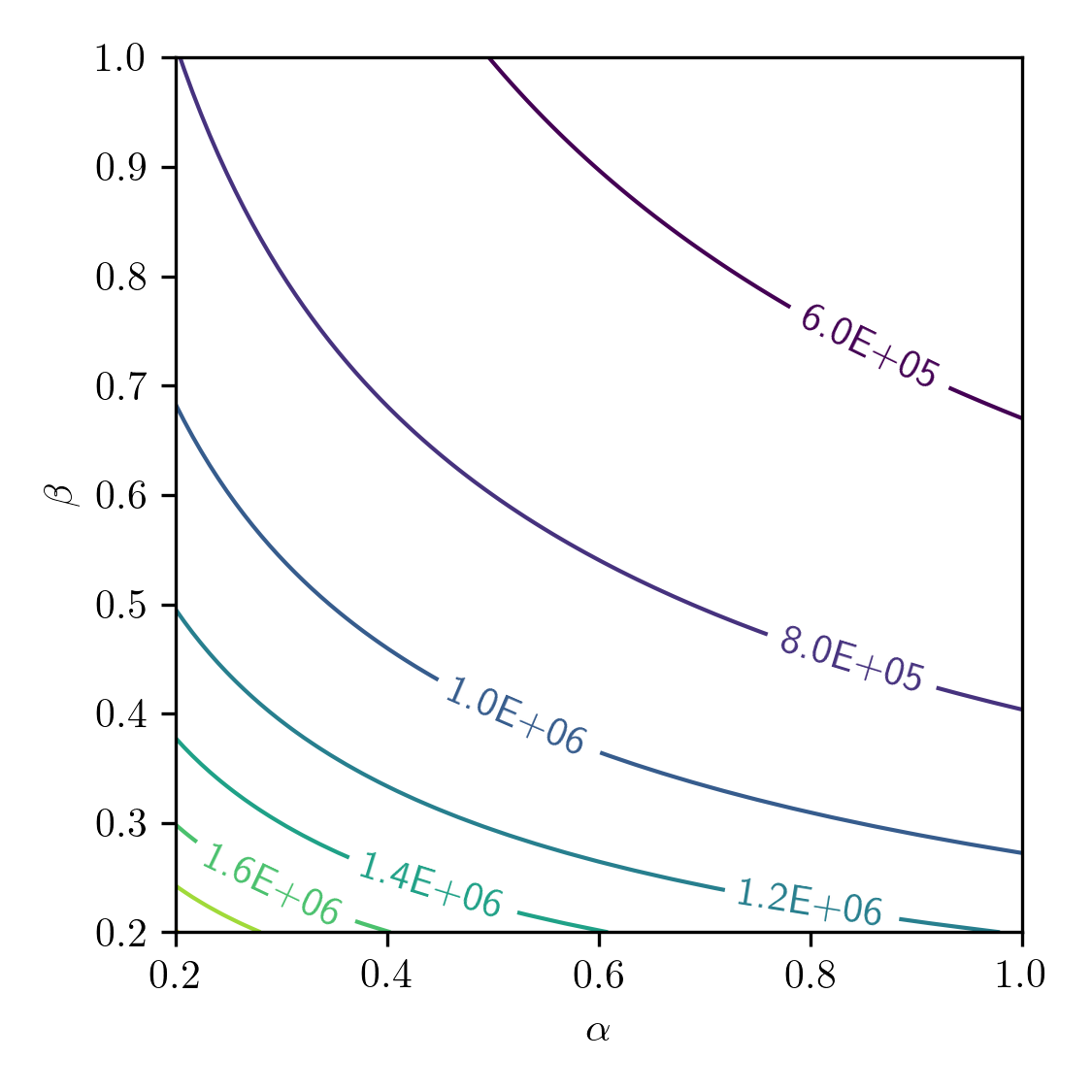}
\caption{\label{fig:rates} Merger rate dependence on $\alpha$ and $\beta$ for a
fixed dark matter fraction ($f = 0.5$) and primordial black hole mass
($M_{\text{BH}} = 1.0 M_\odot$), shown in units of $\text{Gpc}^{-3}
\text{yr}^{-1}$. The expected merger rate strictly increases as either $\alpha$ or
$\beta$ are changed from $1.0$. Similar behavior is observed independent of the
black hole mass or dark matter fraction. This implies that the constraints on
the dark matter fraction that are typically published assuming $\alpha=\beta=1$
are conservative for this model.} \end{figure}

\begin{figure*}%
\centering \subfloat[Dark matter constraint dependence on $\alpha$ and $\beta$]{{\includegraphics[width=.49\textwidth]{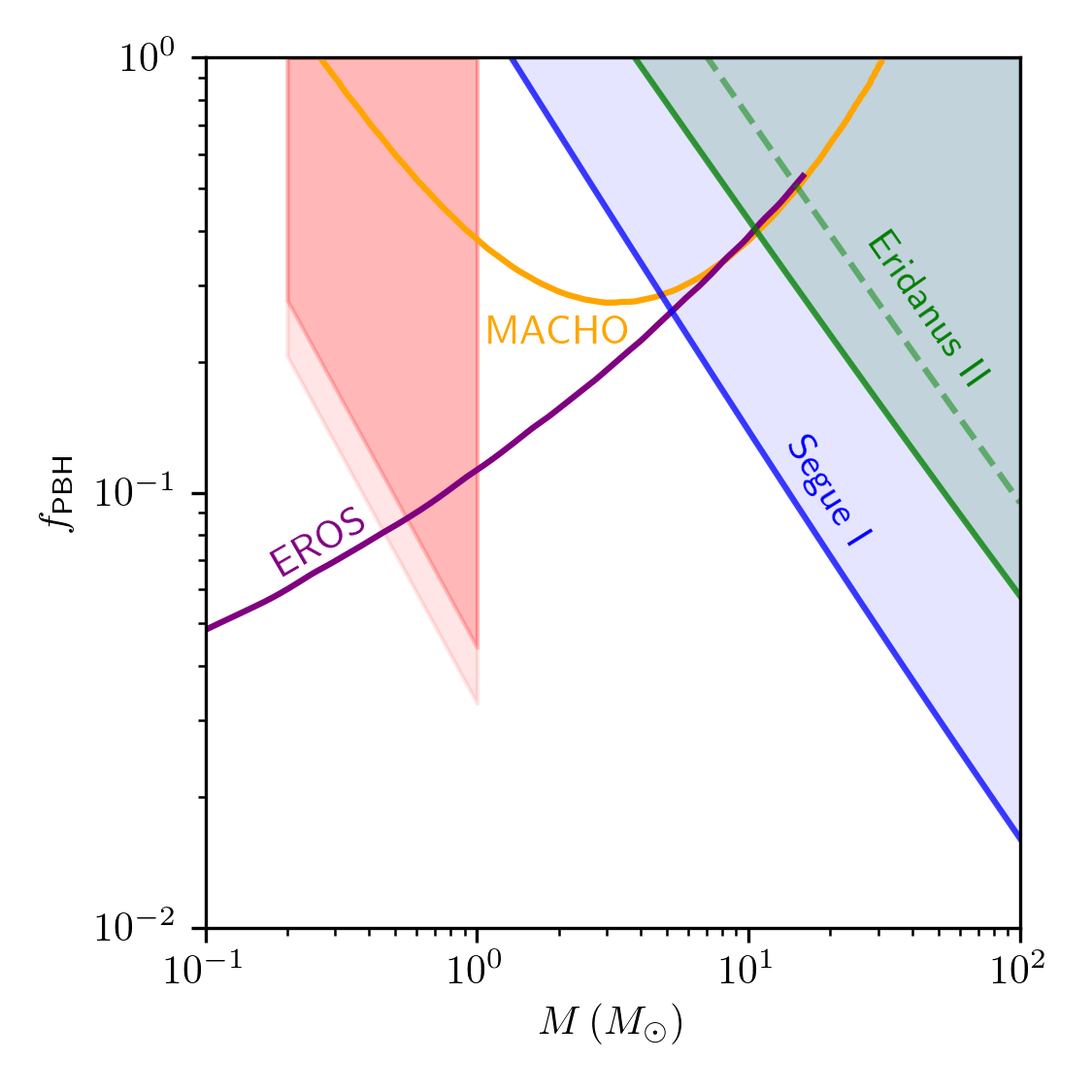} }\label{fig:bounds}}%
\hfill
\centering
\subfloat[Future outlook for LIGO bounds on PBH dark
matter]{{\includegraphics[width=.49\textwidth]{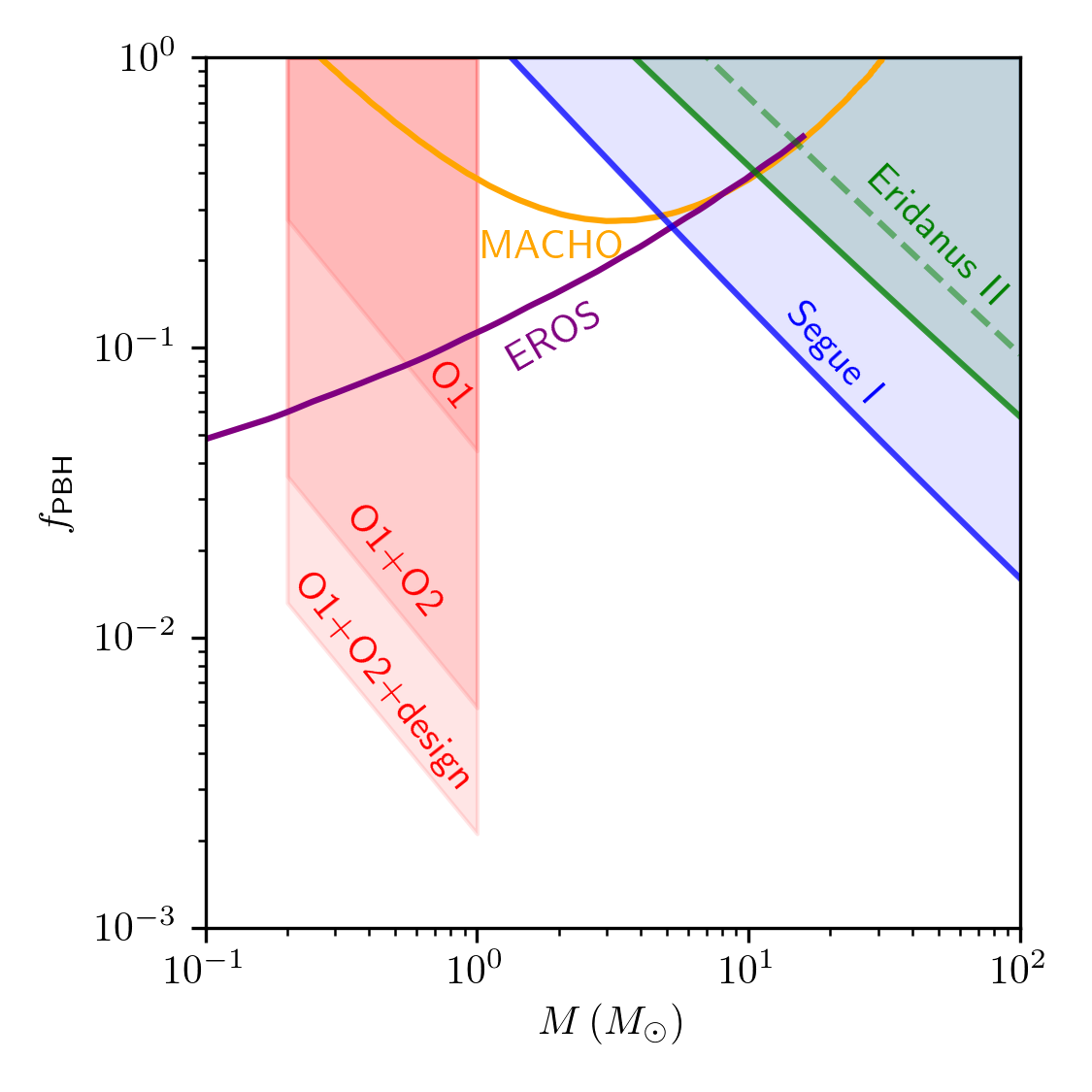}
}\label{fig:future}} \caption{\ref{fig:bounds} Limits on the fraction of dark
matter composed of primordial black holes in a monochromatic distribution.
Shown in purple, yellow, blue, and green are reproductions of the constraints
found in
~\cite{Tisserand:2006zx,Allsman:2000kg,Koushiappas:2017chw,Brandt:2016aco},
respectively. Unlike in ~\cite{O1SSM2018}, the LIGO limits presented here are
based on horizon distance estimates using the power spectra and the loudest
event statistic~\cite{Abadie:2010cg,Biswas:2007ni}. This method is described in
the text. Potential LIGO results shown in red emphasize the small effect
$\alpha$ and $\beta$ have on the constraints. The bottom line shows the limit
for $\alpha = 0.4$, $\beta = 0.8$, while the upper line shows $\alpha = \beta =
1$.~\ref{fig:future} A possible outlook to the future. Shown here are
constraints derived from the same formalism (and assuming continued null
results). We follow the procedure mentioned in the text to approximate the
rates constraints. Here we assume year long runs operating at $40\%$ efficiency
for the O2 and design contributions. LIGO will be able to place percent level
limits on the fraction of dark matter in PBHs after a year of operating at
design sensitivity. The noise curves used for this plot come from the data
release associated with~\cite{Martynov:2016fzi}, specifically the ``Early
high/Mid low'' column for O1, ``Mid high/Late low'' for O2, and ``Design'' for
design.}
\end{figure*}

It is important to consider the robustness of this model and the relative
strictness of the constraints it provides. First, consider the effects of
varying $\alpha$ and $\beta$. Numerical simulations suggest that realistic
values are $\alpha=0.4$, $\beta=0.8$~\cite{Ioka1998}. Though not immediately
evident from the above equation, smaller values of $\alpha$ and $\beta$ lead to
larger \emph{expected} rates and therefore more stringent estimates of the
upper limit of $f_{\text{PBH}}$. The dependence of the expected rate on $\alpha$ and
$\beta$ is shown explicitly in Figure ~\ref{fig:rates}. As $\alpha$ and $\beta$
dip below 1, the expected merger rate increases. It is a simple extension to
approximate how the constraints on $f_{\text{PBH}}$ are affected by variations of
$\alpha$ and $\beta$. We can use the procedure outlined in
~\ref{rates_estimate} to approximate the upper limit on the merger rate, which
we then invert to find limits on $f_{\text{PBH}}$. We present bounds under this
approximation for $\alpha=\beta=1$ and $\alpha=0.4$, $\beta=0.8$ in
Figure~\ref{fig:bounds}. This figure shows a general feature of the model: as either
$\alpha$ or $\beta$ is decreased, the constraint on $f_{\text{PBH}}$ for a given upper
bound on the merger rate becomes tighter. Thus $\alpha=\beta=1$ provides a more
conservative limit on $f_{\text{PBH}}$. 

Of course, allowing $\alpha$, $\beta$ to increase beyond 1 yields looser
constraints. At the time that two PBHs become gravitationally bound to one
another, $\alpha$ describes the ratio between the semi-major axis of the binary
and the initial physical separation of the two PBHs at the moment they become
bound. It is therefore unphysical to expect $\alpha > 1$. $\beta$ helps to
determine the minimum ellipticity of the binary; for $\beta > 1$, the
ellipticity becomes imaginary. $\alpha = \beta = 1$ therefore provides the
\emph{most} conservative rate estimate for this model. 
 
Another important consideration is the sensitivity of this model to errors in
observational measurements of the merger rate. We can propagate errors in rates
measurements through to the dark matter fraction. From our upper limit on the
merger rate estimate, we find that $f_{\text{PBH}} \approx .28$ at $0.2M_\odot$
and $f_{\text{PBH}} \approx .04$ at $1.0 M_\odot$. If we allow for a $50\%$
error in the merger rate estimate that this procedure provides we still find $f
\in (.17, .37 )$ and $f \in (.03, .06 )$ for the respective mass bins, thus
demonstrating that the constraints are relatively insensitive to even large
errors in the upper bound on the merger rate.  

There are several other assumptions made in this model that we do \emph{not}
attempt to quantify, but instead provide a brief qualitative argument on their
effects. First, we have assumed that primordial black holes are uniformly
distributed in space. In reality, we expect PBHs to cluster to some extent
which would change the expected event rate for PBH binary mergers. Clustering
would tend to increase the amount of binary coalescences, however, so the
\emph{expected} event rate would rise and therefore the maximum permissible
fraction, $f_{\text{PBH}}$, would decrease. Therefore a spatially uniform
distribution of PBHs provides a conservative bound on $f_{\text{PBH}}$.  We
also ignore the binary's evolution between formation and coalescence, as well
as the possibility of late-universe binary formation. For a discussion of these
effects, which appear to be sub-dominant (though they also drive the expected
rate up), see~\cite{Ali-Haimoud:2017rtz}. A potentially larger effect comes
from the assumption of a purely monochromatic distribution of black holes.
Though the framework for this formation model has been extended to the unequal
mass case in~\cite{Ioka1998}, we have not considered those effects here.
Finally, we also ignore the effects of spin on binary formation. 

As Advanced LIGO approaches design sensitivity, its horizon distance should
increase by a factor of $2 - 3$~\cite{Aasi:2013wya}. This, coupled with the
more observation, means that LIGO could conceivably have a (cumulative)
sensitive $\langle VT \rangle$ $\mathcal{O}(10)$ times larger than was observed
in ~\cite{O1SSM2018}. Figure ~\ref{fig:future} shows projections for how
continued null results could contribute to constraints on $f_{\text{PBH}}$ for
this mass range. Ground based interferometers have the unique ability to
strengthen bounds in the sub-solar mass regime by systematics independent of
previous microlensing observations
~\cite{Allsman:2000kg,Tisserand:2006zx,Wyrzykowski:2010mh}. This is especially
important in light of recent criticisms~\cite{Hawkins:2015uja} and studies of
the model dependencies of these surveys~\cite{Green:2017qoa}.

\section{Future prospects and discussion}

There are many areas in which sub-solar mass searches can improve on the
suggestions outlined here. The most obvious are extensions to lower masses and
spinning binaries, each of which presents its own challenges. Lower masses
require denser template banks and they persist in LIGO's sensitive band longer.
One possible solution could be to alter the width of the frequency band
considered for different mass bins, thus stitching together a suitable template
bank. Spin is more difficult to incorporate; early tests seem to imply at least
a factor of $10$ more templates would be required for fully spinning binaries.
Examining smaller component spins, such as $\chi_i<0.3$, could remain
computationally feasible and help to mitigate the rapid fall off in sensitivity
that non-spinning banks currently experience for moderate to high spin systems.
We are actively pursuing extensions in these areas.  

More careful PBH population modeling is also a necessity. In particular, a
careful consideration of extended PBH distributions will offer more accurate
and general merger rate predictions. Not only will this allow for more precise
constraints, but it will also be useful in examining the feasibility of
detecting preferred PBH distributions peaked in this mass range.  While this
paper has demonstrated that the model considered typically provides a
conservative estimate of the bounds on $f_{\text{PBH}}$, a more general
formalism will allow testing of different inflationary models.

\section{Acknowledgments}

Funding for this project was provided by the Charles E.  Kaufman Foundation of
The Pittsburgh Foundation.  Computing resources and personnel for this project
were provided by the Pennsylvania State University. We thank the LIGO CBC
working group and John Whelan, Nelson Christensen, and Graham Woan for many
useful questions and comments. We thank Kipp Cannon for suggesting we consider
non-spinning searches in the mass range considered. This document has been
assigned the document number LIGO-P1800231-v3.

\bibliographystyle{unsrt}
\bibliography{references}

\end{document}